\documentstyle[aps,version2,preprint]{revtex}    
\begin{document}

\draft

\begin{title}
Conservation laws and bosonization in integrable Luttinger liquids
\end{title}

\author{J.M.P. Carmelo$^{1,2}$, A.H. Castro Neto$^{2}$,
and D.K. Campbell $^{2}$}
\begin{instit}
$^{1}$ Instituto de Ciencia de Materiales, C.S.I.C.,
Cantoblanco, SP - 28949 Madrid, Spain
\end{instit}
\begin{instit}
$^{2}$ Department of Physics, University of Illinois at
Urbana -- Champaign, \\
1110 West Green Street, Urbana, Illinois 61801-3080
\end{instit}
\receipt{}

\begin{abstract}
We examine and explain the Luttinger-liquid character of models
solvable by the Bethe ansatz by introducing a suitable bosonic
operator algebra. In the case of the Hubbard chain, this involves
two bosonic algebras which apply to {\it all} values of $U$,
electronic density, and magnetization. Only at zero magnetization
does this lead to the usual charge - spin separation. We show
that our ``pseudoparticle'' operator approach clarifies, unifies,
and extends several recent results, including the existence of
independent right and left equations of motion and the concept
of ``pseudoparticle'' (also known as ``Bethe quasiparticle'').
\end{abstract}
\renewcommand{\baselinestretch}{1.656}   


\narrowtext

Based on the similarities between the low-energy spectra of
the Luttinger model and of single-component integrable models
solvable by the Bethe ansatz (BA), Haldane \cite{Haldane}
introduced the concept of ``Luttinger liquid''. However, in
contrast to the case of the Luttinger model \cite{Haldane}, the
study of the BA solvable models did not include an {\it operator}
description of the low-energy physics: the Luttinger-liquid
character of these models relied on the identification
of a structure in the low-energy spectrum provided by the BA.
Further, although there has been considerable progress
in understanding the critical-energy spectrum of multicomponent
integrable systems by combining BA and conformal-field theory
\cite{Frahm}, this approach also does not provide an operator
description of the low-energy problem. Recent studies by Di
Castro and Metzner \cite{Castro} have revealed that instead of
using bosonization \cite{Haldane}, one can construct the
Luttinger-liquid theory for a general ``g-ology'' model in terms of
separate charge and spin conservation for states near the
left and right ``Fermi points''. Again, this construction
was not extended to integrable models solvable by BA and
its interpretation in an operator context was unclear
for these models.

In the present letter we solve the problem of presenting
a general operator approach that establishes the
Luttinger-liquid character of integrable models solvable by
BA. In particular, we use the pseudoparticle-operator
algebra \cite{Neto} to construct a general Luttinger-liquid
theory, which extends the study of Ref. \cite{Castro} to the
case of BA integrable models by establishing explicitly
the separate left-right conservation laws for these
models. Moreover, we introduce the bosonic-operator algebra which
describes the low-energy physics of integrable Luttinger liquids
and justifies the similarities of the energy spectra of these
systems to those of the Luttinger and g-ology models.
We discuss the nature of the pseudoparticles, focusing on the
transformation between electrons and pseudoparticles. This
discussion confirms that the pseudoparticles cannot
be ``removed'' from the many-body system \cite{Carmelo92} and
allows us to write the pseudoparticle generators of the bosonic
excitations explicitly in the electronic basis. More
generally, our results establish a consistent relation
between the Luttinger liquid introduced by Haldane
\cite{Haldane} and the Landau-liquid character of the $U(1)$
sectors of parameter space of integrable Luttinger liquids
\cite{Neto,Carmelo92,Campbell}. Although our present analysis
refers to systems with local interactions, we will demonstrate
explicitly elsewhere that it can be extended to the quantum-chaos
related $1/r^2$ integrable models \cite{Shastry}. The ``Bethe
quasiparticles'' of these models \cite{Shastry} are precisely
the present pseudoparticles, which were first introduced in Refs.
\cite{Carmelo92} (see also references therein) and refer to a large
class of integrable quantum liquids \cite{Neto}.

We consider the particular case of the Hubbard chain
\cite{Lieb} in a external magnetic field $H$ and chemical
potential $\mu$ \cite{Neto,Carmelo92}. This describes
$N_{\uparrow}$ up-spin and $N_{\downarrow}$ down-spin
electrons in a lattice of $N_a$ sites ($N=N_{\uparrow}+
N_{\downarrow}$, $n=N/N_a$, $n_{\sigma }=N_{\sigma }/N_a$,
and $m=n_{\uparrow}-n_{\downarrow}$). We consider densities
$0<n<1$ and spin densities $0<m<n$. This corresponds to the
sector of parameter space with $U(1)\times U(1)$ symmetry
\cite{Neto,Carmelo92}. In this case the low-energy physics
is dominated by a particular class of Hamiltonian eigenstates:
these lowest-weight states (LWS) of both the ``eta''-spin and
spin algebras \cite{Essler} which refer to real rapidities
\cite{Neto,Carmelo92}. We call these states ``LWS I'' to
distinguish them from the LWS associated with complex, non-real,
rapidities, which we call ``LWS II''. Importantly, in the
$U(1)\times U(1)$ sector both the LWS II and the non-LWS
have energy gaps relative to each canonical-ensemble
ground state \cite{Nuno} and do not contribute to the
low-energy physics \cite{Neto}. In the pseudoparticle basis
\cite{Neto,Carmelo92,Campbell,Nuno} one can define a
many-pseudoparticle perturbation theory in which the
non-interacting pseudoparticle ground state is the exact
ground state of the many-electron problem \cite{Neto,Nuno}.
All LWS I can be generated by acting on the vacuum
$|V\rangle$ (zero-electron state) with pseudoparticle operators
\cite{Neto,Nuno}. This operator algebra involves two types
of {\it pseudoparticle} creation (annihilation) operators
$b^{\dag }_{q\alpha }$ ($b_{q\alpha }$) which obey the usual
fermionic algebra \cite{Neto}, $\{b^{\dag }_{q\alpha},
b_{q'\alpha'}\} =\delta_{q,q'}\delta_{\alpha ,\alpha'}$ and
$\{b_{q\alpha},b_{q'\alpha'}\}=0$. Here $\alpha$ refers to the
two pseudoparticle colors $c$ and $s$ \cite{Neto,Carmelo92}.
The discrete pseudomomentum values are $q_j={2\pi\over {N_a}}
I_j^{\alpha }$, where $I_j^{\alpha }$ are {\it consecutive}
integers or half integers. There are $N_{\alpha }^*$ values of
$I_j^{\alpha }$, {\it i.e.} $j=1,...,N_{\alpha }^*$. An LWS I
is specified by the distribution of $N_{\alpha }$ occupied
values, which we call $\alpha $ pseudoparticles, over the
$N_{\alpha }^*$ available values. There are
$N_{\alpha }^*-N_{\alpha }$ corresponding empty values,
which we call $\alpha $ pseudoholes. In the case of the Hubbard
chain, we have that $N_c^*=N_a$, $N_c=N$, $N_s^*=N_{\uparrow}$,
and $N_s=N_{\downarrow}$, {\it i.e.}, the pseudoparticle numbers
are {\it good quantum numbers}. The boundary conditions fix the
numbers $I_j^{\alpha }$: $I_j^c$ are integers (or half integers)
for $N_{\downarrow}$ even (or odd), and $I_j^s$ are integers
(or half integers) for $N_{\uparrow}$ odd (or even)
\cite{Neto,Lieb}. The ground state associated with a
canonical ensemble of $(N_{\uparrow},N_{\downarrow})$ values
[and $(N_c=N_{\uparrow}+N_{\downarrow},N_s=N_{\downarrow}$)]
has the form \cite{Neto,Nuno}

\begin{equation}
|0;N_{\uparrow},N_{\downarrow}\rangle = \prod_{\alpha=c,s}
[\prod_{q=q_{F\alpha }^{(-)}}^{q_{F\alpha }^{(+)}}
b^{\dag }_{q\alpha }]
|V\rangle \, ,
\end{equation}
where when $N_{\alpha }$ is odd (even) and $I_j^{\alpha }$ are
integers (half integers) the pseudo-Fermi points are symmetric
and given by $q_{F\alpha }^{(+)}=-q_{F\alpha }^{(-)}=
{\pi\over {N_a}}[N_{\alpha}-1]$. On the other hand, when
$N_{\alpha }$ is odd (even) and $I_j^{\alpha }$ are half
integers (integers) we have that $q_{F\alpha }^{(+)}=
{\pi\over {N_a}}N_{\alpha }$ and $-q_{F\alpha }^{(-)}
={\pi\over {N_a}}[N_{\alpha }-2]$, or $q_{F\alpha }^{(+)}=
{\pi\over {N_a}}[N_{\alpha }-2]$ and $-q_{F\alpha }^{(-)}=
{\pi\over {N_a}}N_{\alpha }$. In the pseudoparticle basis the
ground state of the many-electron problem is a ``non-interacting''
pseudoparticle ground state of simple Slater-determinant form,
$(1)$. When all $q_{F\alpha }^{(\pm)}$ are symmetric this state
has zero momentum and is non-degenerate, whereas when any number
of these points are nonsymmetric it has finite momentum
and is degenerate \cite{Nuno}. Except for terms of $1/N_a$
order, $q_{Fc }^{(\pm)}=\pm 2k_F=\pm \pi n$ and
$q_{Fs }^{(\pm)}=\pm k_{F\downarrow}=\pm \pi n_{\downarrow}$.

Since the pseudoparticles dominate the physics at
energy scales smaller than the gaps for the LWS II and
non-LWS, it is clearly of interest to construct the
electron-pseudoparticle operator transformation in
this region. The explicit construction of this
transformation for {\it general} BA solvable models is
a very involved unsolved problem. For instance, we
know that at low energies and finite momenta it maps
one-pair electron operators onto multipair pseudoparticle
operators \cite{Campbell}. However, although the one-pair
electronic operator $c^{\dag }_{k'+k\sigma}c_{k'\sigma}$ is,
in general, of multipair pseudoparticle character
($\sigma=\uparrow,\downarrow$ as an operator index and
$\sigma = \pm 1$ otherwise), ${\hat{\rho}}_{\sigma }(k)=
\sum_{k'}c^{\dag }_{k'+k\sigma}c_{k'\sigma}$
is at $k=0$ of one-pair pseudoparticle character.
This result also holds for (nearly) vanishing momenta $k=\pm
{2\pi\over {N_a}}$, as we discuss later. We can write
the pseudoparticle and electronic number operators
${\hat{N}}_{\alpha}={\hat{\rho}}_{\alpha }(0)$ (with
${\hat{\rho}}_{\alpha }(k)=\sum_{q}b^{\dag }_{q+k\alpha}
b_{q\alpha}$) and ${\hat{N}}_{\sigma}={\hat{\rho}}_{\sigma }(0)$,
respectively, in the electronic and pseudoparticle basis
as ${\hat{N}}_{\alpha} = \sum_{\sigma}G_{\alpha\sigma}
{\hat{N}}_{\sigma}$ and ${\hat{N}}_{\sigma} = \sum_{\sigma}
G^{-1}_{\sigma\alpha}{\hat{N}}_{\alpha}$, respectively, where
${\rm\bf G}$ and ${\rm\bf G}^{-1}$ are the electron -
pseudoparticle and pseudoparticle - electron charge
matrices, respectively. Their entries are $G_{c\uparrow}=
G_{c\downarrow}=G_{s\downarrow}=1$, $G_{s\uparrow}=0$
and $G^{-1}_{\uparrow c}=-G^{-1}_{\uparrow s}=
G^{-1}_{\downarrow s}=1$, $G^{-1}_{\downarrow c}=0$.
The simple form of these expressions follows from the
conservation of the number of $\alpha$ pseudoparticles,
$N_{\alpha}$, so that $N_{\alpha}=N_{\gamma (\alpha)}$,
where $\gamma (c)=\rho$ and $\gamma (s)=\downarrow$.
However, that ${\hat{\rho}}_{\alpha }(0)=
{\hat{\rho}}_{\gamma (\alpha)}(0)$  does {\it not}
imply that ${\hat{\rho}}_{\alpha }(k)=
{\hat{\rho}}_{\gamma (\alpha)}(k)$ for $k>0$; we shall
return to this important point later.

The non-perturbative character of the usual electronic
basis -- {\it e.g.}, the absence of a coherent $\delta$-function
peak in the one-electron spectral function --
is reflected in the properties of the
electron - pseudoparticle transformation. Let us
use the changes in the quantum numbers $I_j^{\alpha}$
between ground states $(1)$ differing in the electronic or
pseudoparticle numbers by $\pm 1$ \cite{Nuno} to describe
one electron of lowest-energy in terms of pseudoparticles and
to show that the latter {\it cannot} be removed from the
many-electron system \cite{Carmelo92}. For simplicity, we
choose both numbers $N_{\sigma}$ to be odd in the starting
state $(1)$ which we call $\cal A$ (our results are independent
of this choice). This state has zero momentum and is non-degenerate.
To define the up-spin electron in terms of
pseudoparticles we compare the state  $\cal A$ with the
ground state $|0;N_{\uparrow}\pm 1,N_{\downarrow}\rangle$.
This has pseudoparticle numbers $N_c=N\pm 1$ and $N_s=
N_{\downarrow}$ (state $\cal B$). The changes in the electronic
and pseudoparticle numbers seem to indicate that removing or
adding one up-spin electron is equivalent to removing or
adding one $c$ pseudoparticle. However, this is not true.
Looking carefully at the changes in the pseudoparticle quantum
numbers $I_j^{\alpha }$ from  $\cal A$ to $\cal B$, we find
that within the many-body system the up-spin electron
consists of {\it two} excitations which {\it cannot} be decomposed:
1) an excitation removing or adding one $c$ pseudoparticle from one
of the pseudo-Fermi points, with momentum
$q_{Fc}^{(\pm)}=\pm 2k_F$; and 2) a collective
excitation involving {\it all} $s$ pseudoparticles
which shifts their pseudomomenta by $\mp {\pi\over {N_a}}$.
This shift results from the change of the quantum
numbers $I^s_j$ from integers to half integers. Although
each pseudomomentum shift, $\mp {\pi\over {N_a}}$, is vanishing
small, adding all contributions from the $s$ pseudo-Fermi
sea gives $\mp k_{F\downarrow }$. The total momentum
of the up-spin electron is $P = \pm [2k_{F} -k_{F\downarrow}]=
\pm k_{F\uparrow}$ (here and below we omit $1/N_a$ corrections
to $P$). Since excitations 1) and 2)
cannot be decomposed, we cannot remove or add
one $c$ pseudoparticle without making a collective
excitation involving all $s$ pseudoparticles.
It is in this sense that we say that the $c$
pseudoparticle cannot be removed from the
many-body system and cannot exist as a free entity.
A similar analysis shows that the down-spin electron
is a collective pseudoparticle object made out of
one $c$ pseudoparticle, one $s$ pseudoparticle, and a
pseudomomentum shift of all remaining $c$ pseudoparticles.
Its momentum is $P = \pm 2k_{F}\mp 2k_{F}\pm k_{F\downarrow}=
\pm k_{F\downarrow}$. Both the up- and down-spin electrons
can be removed from and added to the system, but within
the system they do not refer to Hamiltonian eigenstates
and have zero lifetime: when added to the system
they decay immediately into collective pseudoparticle
excitations. Similarly, the $s$ pseudoparticle cannot
exist as a free entity out of the many-electron system:
the ground state with numbers $N_c=N$ and
$N_s=N_{\downarrow}\pm 1$, $|0;N_{\uparrow}\mp 1,
N_{\downarrow}\pm 1\rangle$, is, relative to the state
$\cal A$, nothing but an electronic spin flip. This is a
collective pseudoparticle object made out of one $s$
pseudoparticle, a pseudomomentum shift of all remaining $s$
pseudoparticles, and a pseudomomentum shift of all $c$
pseudoparticles. These three excitations {\it cannot} be
decomposed.

Let $\cal {H}_I$ be the Hilbert space spanned by the LWS
I. In $\cal {H}_I$ and normal-ordered relative to $(1)$,
the Hubbard Hamiltonian has an infinite number of terms which
correspond to increasing ``order'' of pseudoparticle scattering
\cite{Neto}: formally, $:\hat{H}:=\sum_{i=1}^{\infty}
\hat{H}^{(i)}$, where the first two terms are
$\hat{H}^{(1)}=\sum_{q,\alpha}\epsilon_{\alpha}(q)
:\hat{N}_{\alpha}(q):$ and $\hat{H}^{(2)}={1\over
{N_a}}\sum_{q,\alpha} \sum_{q',\alpha'}{1\over 2}
f_{\alpha\alpha'}(q,q'):\hat{N}_{\alpha}(q):
:\hat{N}_{\alpha'}(q'):$. The expressions for the pseudoparticle
bands, $\epsilon_{\alpha}(q)$, and of the ``Landau''
$f$ functions, $f_{\alpha\alpha'}(q,q')$, are given in
Ref. \cite{Carmelo92}. The latter involve the velocities
$v_{\alpha}(q) = {d\epsilon_{\alpha}(q)\over {dq}}$ and the
two-pseudoparticle forward-scattering phase shifts
$\Phi_{\alpha\alpha'}(q,q')$. These are defined in the second
paper of Ref. \cite{Carmelo92}. In particular, the velocities
$v_{\alpha}\equiv v_{\alpha}(q_{F\alpha}^{(+)})$ and the
parameters $\xi_{\alpha\alpha '}^j=\delta_{\alpha\alpha '}+
\Phi_{\alpha\alpha '}(q_{F\alpha}^{(+)},q_{F\alpha '}^{(+)})+
(-1)^j\Phi_{\alpha\alpha '}(q_{F\alpha}^{(+)},q_{F\alpha '}^{(-)})$,
with $j=0,1$, play a determining role at the critical point.
($\xi_{\alpha\alpha '}^1$ are the entries of the transpose
of the dressed-charge matrix \cite{Frahm}.) An essential point
is that at constant values of the electron and pseudoparticle
numbers the electron - pseudoparticle transformation {\it does
not} mix left and right electronic operators, {\it i.e.}, $\iota =sgn
(k)1=\pm 1$ electronic operators are made out of $\iota =sgn
(q)1=\pm 1$ pseudoparticle operators only, $\iota $ defining
the right ($\iota=1$) and left ($\iota=-1$) movers.
Measuring the electronic momentum $k$ and
pseudomomentum $q$ from the $U=0$ Fermi points
$k_{F\sigma}^{(\pm)}=\pm \pi n_{\sigma}$ and  pseudo-Fermi
points $q_{F\alpha}^{(\pm)}$, respectively, adds the index
$\iota$ to the electronic and pseudoparticle operators.
The new momentum $\tilde{k}$ and pseudomomentum $\tilde{q}$ are
such that $\tilde{k} =k-k_{F\sigma}^{(\pm)}$ and
$\tilde{q}=q-q_{F\alpha}^{(\pm)}$, respectively, for $\iota=\pm 1$.
For instance, ${\hat{\rho}}_{\sigma\iota }(k)=\sum_{\tilde{k}}
c^{\dag }_{\tilde{k}+k\sigma\iota}c_{\tilde{k}\sigma\iota}$ and
${\hat{\rho}}_{\alpha\iota }(k)=\sum_{\tilde{q}}b^{\dag
}_{\tilde{q}+k\alpha\iota}b_{\tilde{q}\alpha\iota}$. The perturbative
character of the pseudoparticle basis \cite{Neto} implies
that the above pseudoparticle Hamiltonian can be used
as starting point for the construction of a critical-point
Hamiltonian. This proceeds by linearizing the pseudoparticle bands
$\epsilon_{\alpha}(q)$ around the pseudo-Fermi points. In addition,
one considers the terms associated with two-pseudoparticle
scattering near the pseudo-Fermi points {\it only} and
replaces the full $f$ functions by $f_{\alpha\alpha'}^{1}=
f_{\alpha\alpha'}(q_{F\alpha}^{(\pm)},
q_{F\alpha'}^{(\pm)})$ and $f_{\alpha\alpha'}^{-1}=
f_{\alpha\alpha'}(q_{F\alpha}^{(\pm)},q_{F\alpha'}^{(\mp)})$.
(The expressions of $f_{\alpha\alpha'}^{\pm 1}$ involve the
velocities $v_{\alpha}$ and parameters $\xi_{\alpha\alpha '}^j$
only \cite{Neto}.) The critical-point Hamiltonian can be written as
$:\hat{{\cal H}}:={\hat{{\cal H}}}_0 + {\hat{{\cal H}}}_2
+ {\hat{{\cal H}}}_4$, where ${\hat{{\cal H}}}_0 =
\sum_{\alpha ,\iota ,\tilde{q}}
\iota v_{\alpha}\tilde{q}  :\hat{N}_{\alpha,\iota}(\tilde{q}):$
and ${\hat{{\cal H}}}_2 + {\hat{{\cal H}}}_4 =
{2\over {N_a}}\sum_{\alpha,\alpha',\iota,\iota',k}
[g_2^{\alpha\alpha '}(k)\delta_{\iota ,-\iota '} +
g_4^{\alpha\alpha '}(k)\delta_{\iota ,\iota '}]
:{\hat{\rho}}_{\alpha ,\iota}(k):
:{\hat{\rho}}_{\alpha ',\iota'}(-k):$. Here
${\hat{N}}_{\alpha\iota}(\tilde{q})=
b^{\dag }_{\tilde{q}\alpha\iota}b_{\tilde{q}\alpha\iota}$
and the couplings read $g_2^{\alpha\alpha '}(k)=
f_{\alpha\alpha'}^{-1}\delta_{k,0}$ and $g_4^{\alpha\alpha '}(k)=
f_{\alpha\alpha'}^{1}\delta_{k,0}$. It follows that in the
pseudoparticle basis the suitable critical-point Hamiltonian is a
g-ology model of the type studied in Ref. \cite{Castro}
with exotic $k=0$ forward-scattering couplings, $\sigma$ replaced by
$\alpha$, and the electronic operators by pseudoparticle operators.
We emphasize that the existence of only $k=0$ pseudoparticle
scattering allows the ground-state pseudomomentum distribution,
$\langle \hat{N}_{\alpha}(q)\rangle $, to be of non-interacting
character, {\it i.e.} equal to 1 inside and 0 outside the
pseudo-Fermi surface, respectively \cite{Neto,Carmelo92,Nuno}.
The absence of the $g_1$ and $g_3$ terms implies that the
$\alpha,\iota$ pseudoparticle number operators,
$\hat{N}_{\alpha,\iota}=\sum_{\tilde{q}}
\hat{N}_{\alpha,\iota}(\tilde{q})$, such that
$\hat{N}_{\alpha,\iota}=\hat{N}_{\gamma (\alpha),\iota}$,
are good quantum numbers, {\it i.e.} there are {\it
separate} right and left conservation laws. This is a
generalization of the results of Ref. \cite{Castro} with
the Fermi points replaced by the pseudo-Fermi points.
We emphasize that, in the case of a single-component
model, as well as for the anisotropic $S=1/2$ Heisenberg chain
\cite{Haldane} and the Hubbard chain at half filling
\cite{Frahm,Carmelo92}, we can omit the index $\alpha$, so
that the critical Hamiltonian can be rewritten as
$:\hat{{\cal H}}:={\hat{{\cal H}}}_0+\hat{V}$, with
$\hat{V}={\pi\over {N_a}}\sum_{\iota ,\iota ',k}
[V_1(k)\delta_{\iota ,\iota '}+V_2(k)\delta_{\iota ,-\iota
'}]:{\hat{\rho}}_{\iota}(k)::{\hat{\rho}}_{\iota'}(-k):$.
Here ${\hat{\rho}}_{\iota}(k)=\sum_{\tilde{q}}b^{\dag
}_{\tilde{q}+k\iota}b_{\tilde{q}\iota}$,
$V_1(k)={f^{1}\over {2\pi}}\delta_{k,0}$,
and $V_2(k)={f^{-1}\over {2\pi}}\delta_{k,0}$.
Therefore, in {\it single-component} integrable systems,
$:\hat{{\cal H}}:$ is a Luttinger model with exotic $k=0$
forward-scattering potentials. This universal form of
the critical-point Hamiltonian in the pseudoparticle-operator
basis justifies the Luttinger-liquid character of integrable
models by BA.

The separate right and left conservation laws provide
the Luttinger-liquid parameters through equations
of motions \cite{Castro}. Let $\vartheta $ be any conserved
quantity such as charge ($\vartheta =\rho$), spin
($\vartheta =\sigma_z$), or spin projection
($\vartheta =\sigma$). Then ${\hat{N}}_{\vartheta\iota}$
can be written as ${\hat{N}}_{\vartheta\iota}=\sum_{\alpha}
k_{\vartheta\alpha}{\hat{N}}_{\gamma (\alpha)\iota}$,
where the integers $k_{\vartheta\alpha}$ are $k_{\rho
c}=k_{\sigma_{z} c}=k_{\uparrow c}=1$, $k_{\downarrow c}=0$,
$k_{\rho s}=0$, $k_{\sigma_{z} s}=-2$, $k_{\uparrow s}=-1$,
and $k_{\downarrow s}=1$ (note that
$k_{\sigma\alpha}=G^{-1}_{\sigma\alpha}$
and $k_{\alpha\alpha '}=\delta_{\alpha ,\alpha '}$).
The operator ${\hat{\rho}}_{\gamma(\alpha)\iota }(k)$
can be written as ${\hat{\rho}}_{\gamma(\alpha)\iota }(k)=
\sum_{\sigma}G_{\alpha\sigma}{\hat{\rho}}_{\sigma\iota }(k)$.
(Although ${\hat{\rho}}_{\gamma(\alpha)\iota }(0)=
{\hat{\rho}}_{\alpha\iota }(0)$,
${\hat{\rho}}_{\gamma(\alpha)\iota
}(k)\neq {\hat{\rho}}_{\alpha\iota }(k)$ for $k\neq 0$,
as we find below.) Let us consider the general operator
${\hat{\rho}}_{\vartheta\iota }(k)=\sum_{\alpha}k_{\vartheta\alpha}
{\hat{\rho}}_{\gamma(\alpha)\iota }(k)$. It is useful to consider
the combinations ${\hat{\rho}}_{\vartheta }^{(\pm)}(k)=
{\hat{\rho}}_{\vartheta 1}(k) \pm {\hat{\rho}}_{\vartheta -1}(k)$
(note that ${\hat{\rho}}_{\vartheta }^{(+)}(k)=
{\hat{\rho}}_{\vartheta }(k)$). Since the commutator
$[{\hat{\rho}}_{\vartheta }^{(\pm)}(k,t),
{\hat{{\cal H}}}_2+{\hat{{\cal H}}}_4]=0$ for $k>0$ and
$[{\hat{\rho}}_{\vartheta }^{(\pm)}(k,t),:\hat{{\cal H}}:]$ is
proportional to $k$ at $k=0$, the interesting quantity associated
with the equation of motion for
${\hat{\rho}}_{\vartheta }^{(\pm)}(k,t)$ \cite{Castro} is
the following ratio, which we evaluate using
the forms of the coupling constants of $:\hat{{\cal H}}:$

\begin{equation}
{i\partial_t {\hat{\rho}}_{\vartheta }^{(\pm)}(k,t)\over
k}|_{k=0} = {[{\hat{\rho}}_{\vartheta }^{(\pm)}(k,t),
:\hat{{\cal H}}:]\over k}|_{k=0} ={\cal
V}_{\vartheta}^{(\mp)}{\hat{\rho}}_{\vartheta
}^{(\mp)}(0,t)
\end{equation}
where ${\cal V}_{\vartheta}^{(-)} =
\sum_{\alpha ,\alpha'}k_{\vartheta\alpha}k_{\vartheta\alpha'}
[v_{\alpha}\delta_{\alpha ,\alpha '} + {[f_{\alpha\alpha '}^{1}-
f_{\alpha\alpha '}^{-1}]
\over {2\pi}}] = \sum_{\alpha}v_{\alpha}
[\sum_{\alpha'}k_{\vartheta\alpha '}\xi_{\alpha\alpha '}^1]^2$
and ${\cal V}_{\vartheta}^{(+)} = 1/
\{\sum_{\alpha ,\alpha'}{k_{\vartheta\alpha}k_{\vartheta\alpha'}
\over {v_{\alpha}v{\alpha '}}}
[v_{\alpha}\delta_{\alpha ,\alpha '} - {[A_{\alpha\alpha '}^{1}+
A_{\alpha\alpha '}^{-1}]
\over {2\pi}}]\} = 1/\{\sum_{\alpha}{1\over {v_{\alpha}}}
[\sum_{\alpha'}k_{\vartheta\alpha '}\xi_{\alpha\alpha '}^1]^2\}$.
Here $A_{\alpha\alpha'}^{1}=A_{\alpha\alpha'
}(q_{F\alpha}^{(\pm)}, q_{F\alpha'}^{(\pm)})$ and
$A_{\alpha\alpha'}^{-1}= A_{\alpha\alpha'}(q_{F\alpha}^{(\pm)},
q_{F\alpha'}^{(\mp)})$, where $A_{\alpha\alpha'}(q,q')$ are
the scattering amplitudes given by Eqs. $(83)-(85)$ of the
third paper of Ref. \cite{Carmelo92}. The velocities
${\cal V}_{\vartheta}^{(+)}$ and ${\cal V}_{\vartheta}^{(-)}$ determine
the $\vartheta$ susceptibility, $K^{\vartheta }=1/[\pi{\cal
V}_{\vartheta}^{(+)}]$, and the coherent part of the
$\vartheta $ conductivity spectrum, ${\cal V}_{\vartheta}^{(-)}
\delta (\omega )$, respectively \cite{Castro}. This agrees with the
studies of Ref. \cite{Carmelo92} for charge and
spin and with the conductivity expressions of Ref. \cite{Neto}.
For single-component systems there is only one choice for
$\vartheta$ and ${\cal V}^{(-)}=v[\xi^1]^2$
and ${\cal V}^{(+)}=v[\xi^0]^2=v/[\xi^1]^2$, in agreement
with Ref. \cite{Haldane}. The ${\cal V}_{\vartheta}^{(\pm)}$
are the expressions for the $\vartheta$ conserved quantities
of integrable {\it multicomponent} Luttinger liquids. Equation
$(2)$ involves the commutator of the pseudoparticle-Hamiltonian
$:\hat{{\cal H}}:$ with an electronic operator and, therefore,
the velocities ${\cal V}_{\vartheta}^{(\pm)}$ do not have the
same simple form as for the g-ology model of Ref. \cite{Castro}.
Importantly, except for single-component integrable models,
${\cal V}_{\vartheta}^{(+)}$ {\it does not} equal the
expression of ${\cal V}_{\vartheta}^{(-)}$
with $f_{\alpha\alpha '}^{1}-f_{\alpha\alpha '}^{-1}$
replaced by $f_{\alpha\alpha '}^{1}+f_{\alpha\alpha '}^{-1}$.

The bosonization of $:\hat{{\cal H}}:$ is straightforward
and refers to the non-interacting pseudoparticle
ground state $(1)$. We find that $[\hat{\rho}_{\alpha\iota}(k),
\hat{\rho}_{\alpha'\iota'}(-k')]=\delta_{\alpha,\alpha'}
\delta_{\iota ,\iota'}\delta_{k,k'}(\iota
k{N_a\over {2\pi}})$ and the $\alpha$ bosonic
operators are given by

\begin{equation}
a^{\dagger}_{k\alpha} = \sqrt{{2\pi\over
{N_a|k |}}} \sum_{\iota} \theta (\iota k)
\hat{\rho}_{\alpha\iota}(k)  \, ,
\end{equation}
for $k>0$. Our bosonization reproduces the results of conformal-field
theory \cite{Frahm}: the bosons $(3)$ refer to the tower excitations
of Ref. \cite{Neto}, whereas the HWS \cite{Neto} of the Virasoro Algebras
\cite{Frahm}, which introduce the anomalous dimensions,
are the current and ``charge'' excitations \cite{Haldane}. The
low-energy separation refers to the colors $\alpha $
for all parameter space.

To complete our analysis we should show how to express these
bosonic generators $\hat{\rho}_{\alpha\iota}(k)$ in the electronic
basis. This will allow a definition of the $\alpha $ low-energy
separation in terms of usual electronic operators. Let
us consider the reduced Hilbert space spanned by the
Hamiltonian eigenstates of momentum $k=\iota {2\pi\over {N_a}}$
relative to the ground state $(1)$, which we shall
henceforth denote by  $|0\rangle $, so that we
can denote the Hamiltonian eigenstates by $|\alpha\iota\rangle
=\hat{\rho}_{\alpha\iota}(k)|0\rangle $ with
$k=\iota {2\pi\over {N_a}}$. These states are orthogonal
to $(1)$ and have a single pseudohole at one of
the pseudo-Fermi points. In the reduced Hilbert space they
constitute a complete orthonormal basis, so that
$\langle\alpha\iota|\alpha '\iota '\rangle =\delta_{\alpha
,\alpha '}\delta_{\iota ,\iota '}$ and
$\sum_{\alpha ,\iota}|\alpha\iota\rangle\langle\alpha\iota|=
\openone$. Based on the results of Ref. \cite{Carmelo92}
we find that the one-pair electronic operator
$\hat{\rho}_{\sigma\iota}(k)$ at $k=\iota{2\pi\over {N_a}}$ is
of one-pair pseudoparticle character and projects
the ground state $(1)$ in the above reduced Hilbert space,
{\it i.e.}, $\hat{\rho}_{\sigma\iota}(k)|0\rangle\langle 0|=
\sum_{\alpha ,\iota}{\cal U}^{-1}_{\alpha\sigma}
|\alpha\iota\rangle\langle 0|$, where
${\cal U}^{-1}_{\alpha\sigma}=\langle\alpha\iota|
\hat{\rho}_{\sigma\iota}(k)|0\rangle$: at $k=\iota{2\pi\over {N_a}}$
we have that $\hat{\rho}_{\sigma\iota}(k)=\sum_{\alpha ,\iota}{\cal
U}^{-1}_{\alpha\sigma}\hat{\rho}_{\alpha\iota}(k)$ where we denote
${\rm\bf {\cal U}}^{-1}$ by {\it pseudoparticle - electron matrix}.
Introducing the ``electronic'' states $|\sigma\iota\rangle=
\hat{\rho}_{\sigma\iota}(k)|0\rangle $ with $k=\iota{2\pi\over {N_a}}$,
and applying the methods of the third paper of Ref. \cite{Carmelo92}
we find ${\cal U}^{-1}_{\alpha\sigma}=\langle\alpha\iota
|\sigma\iota\rangle=\sum_{\alpha'}G^{-1}_{\sigma\alpha '}
\xi^{1}_{\alpha\alpha'}$. However, here we are interested in
the inverse problem, {\it i.e.}, writing the operators
$\hat{\rho}_{\alpha\iota}(k)$ in the electronic basis.
Fortunately, the $k=\iota{2\pi\over {N_a}}$
states $|\sigma\iota\rangle$ constitute a complete (but
non-orthonormal) basis in the reduced Hilbert space.
This follows from the fact that $\det {\rm\bf {\cal U}}^{-1}>0$.
Therefore, we can invert the matrix ${\rm\bf {\cal U}}^{-1}$
to find

\begin{equation}
\hat{\rho}_{\alpha\iota}(k) =
\sum_{\sigma}{\cal U}_{\sigma\alpha} \hat{\rho}_{\sigma\iota}(k) \, ,
\hspace{1cm} k=\iota {2\pi\over {N_a}} \, ,
\end{equation}
where the electron - pseudoparticle matrix
${\rm\bf {\cal U}}$ is such that ${\cal U}_{\sigma\alpha}=
\sum_{\alpha'}G_{\alpha'\sigma}\xi^{0}_{\alpha\alpha'}$.
In the electronic basis and at $k=\iota {2\pi\over {N_a}}$
the operators $(3)$ read $a^{\dagger}_{k\alpha}=
\sqrt{{2\pi\over {N_a|k |}}} \sum_{\sigma ,\iota} \theta
(\iota k) {\cal U}_{\sigma\alpha}\hat{\rho}_{\sigma\iota}(k)$.
We emphasize that the one-pair electronic operator
$\hat{\rho}_{\sigma\iota}(k)$ does not change the numbers
$N_{\sigma }$ and, therefore, the pseudoparticle-collective
part of the two associated one-electron excitations cancels.
Equation $(4)$ reveals that the generators of the $\alpha$ bosonic
(tower) excitations are an interaction-dependent mixture of up-spin
and down-spin one-pair electronic operators: in the
$U(1)\otimes U(1)$ sector {\it all} the $2\times 2$
matrix elements ${\cal U}_{\sigma\alpha}$ are
non-vanishing and interaction dependent. This
shows the exotic character of the $\alpha$ low-energy
separation, which is different from the {\it dynamical
separation} studied in Ref. \cite{Neto}: this latter effect
refers to the transport masses which, due to the pure $k=0$
forward-scattering character of $:\hat{{\cal H}}:$, only ``see''
the $k=0$ $\alpha$ operators $\hat{\rho}_{\alpha\iota}(0)
=\hat{\rho}_{\gamma (\alpha)\iota}(0)$. This leads to a
{\it dynamical} charge - down-spin separation
\cite{Neto}. On the other hand, the $k>0$ $\alpha$
generators of $(3)$ {\it are not}
the operators $\hat{\rho}_{\gamma (\alpha)\iota}(k)$.
Instead, they are the exotic mixture of the electronic
operators $\hat{\rho}_{\sigma\iota}(k)$, $(4)$.
However, from the usual $m=0$ and (or) $n=1$ pictures
\cite{Haldane,Castro,Emery}, we expect $c$ and
$s$ to become charge and spin, respectively, in the limit
$m\rightarrow 0$, and $c$ to become charge in the limit
$n\rightarrow 1$. This is confirmed by Table 1 where
we show some limiting forms of $(4)$. A more detailed
study of expression $(4)$ will be  presented elsewhere.

This work was supported by C.S.I.C. (Spain) and the University
of Illinois. We thank F. Guinea for stimulating discussions.
A. H. C. N. thanks CNPq (Brazil) for a scholarship.


\newpage
\centerline{TABLE}
\vspace{2.5cm}
\narrowtext
\begin{tabbing}
\sl \hspace{1.3cm} \= \sl $(a)\, U\rightarrow 0$ \hspace{0.3cm} \=
\sl $(b)\, U\rightarrow 0$ \hspace{0.3cm} \= \sl $(c)\,
n\rightarrow 0$ \hspace{0.3cm} \= \sl $(d)\, m\rightarrow 0$
\hspace{0.3cm} \= \sl $(e)\, n_{\downarrow}\rightarrow 0$
\hspace{0.7cm}\= \sl $(f)\, n\rightarrow 1$
\\ $\hat{\rho}_{c\iota}(k)$
\> $\hat{\rho}_{\uparrow\iota}(k)$
\> ${\hat{\rho}_{\rho\iota}(k)\over{\sqrt{2}}}$
\> $\hat{\rho}_{\rho\iota}(k)$
\> $\xi^0_{cc}\hat{\rho}_{\rho\iota}(k)$
\> $\sum_{\sigma} {\cal U}^{-1}_{\sigma c}
   \hat{\rho}_{\sigma\iota}(k)$
\> $\hat{\rho}_{\rho\iota}(k)$\\
$\hat{\rho}_{s\iota}(k)$
\> $\hat{\rho}_{\downarrow\iota}(k)$
\> $-{\hat{\rho}_{\sigma_z\iota}(k)\over {\sqrt{2}}}$
\> $\hat{\rho}_{\downarrow\iota}(k)$
\> $-{\hat{\rho}_{\sigma_z\iota}(k)\over {\sqrt{2}}}$
\> $\hat{\rho}_{\downarrow\iota}(k)$
\> $\sum_{\sigma} {\cal U}^{-1}_{\sigma s}
   \hat{\rho}_{\sigma\iota}(k)$
\label{table1}
\end{tabbing}
\vspace{0.5cm}
TABLE 1 -- Limiting values of the $c$ and $s$ generators
at $k=\iota {2\pi\over {N_a}}$ and for (a) $U\rightarrow 0$
when $0<n<1$ and $n_{\uparrow}
>n_{\downarrow}$; (b) $U\rightarrow 0$ when $0<n<1$ and
$n_{\uparrow}=n_{\downarrow}$; (c) $n\rightarrow 0$ when
$n_{\uparrow}>n_{\downarrow}$ and $U>0$; (d) $m\rightarrow 0$
when $0<n<1$ and $U>0$; (e) $n_{\downarrow}\rightarrow 0$
when $n_{\downarrow}<n_{\uparrow}$ (here ${\cal U}^{-1}_{\uparrow c}
=1$); and (f) $n\rightarrow 1$ when $U>0$.

\end{document}